\documentclass[%
 reprint,
superscriptaddress,
 amsmath,amssymb,
 aps,
]{revtex4-2}

\usepackage{graphicx}
\usepackage{dcolumn}
\usepackage{bm}
\usepackage{xcolor}
\usepackage{color}
\usepackage[utf8x]{inputenc}

\newcommand{\bfS}{{\bf S}}
\newcommand{\bfs}{{\bf s}}

\newcommand{\vecs}{\vec{s}}
\newcommand{\hatsigma}{\hat{\sigma}}

\usepackage[normalem]{ulem}

\begin{document}
\title{
Self-Supervised Ensemble Learning:
A Universal Method for Phase Transition Classification of Many-Body Systems
}

\author{Chi-Ting Ho}
\affiliation{Physics Department, National Tsing Hua University, Hsinchu 30013, Taiwan}
\affiliation{Center for Theory and Computation, National Tsing Hua University, Hsinchu 30013, Taiwan}

\author{Daw-Wei Wang}
\affiliation{Physics Department, National Tsing Hua University, Hsinchu 30013, Taiwan}
\affiliation{Center for Theory and Computation, National Tsing Hua University, Hsinchu 30013, Taiwan}
\affiliation{Center for Quantum Technology, National Tsing Hua University, Hsinchu 30013, Taiwan}
\affiliation{Physics Division, National Center for Theoretical Sciences, Taipei 10617, Taiwan}

\begin{abstract}
We develop a self-supervised ensemble learning (SSEL) method to accurately classify distinct types of phase transitions by analyzing the fluctuation properties of machine learning outputs. Employing the 2D Potts model and the 2D Clock model as benchmarks, we demonstrate the capability of SSEL in discerning first-order, second-order, and Berezinskii-Kosterlitz-Thouless transitions, using in-situ spin configurations as the input features. Furthermore, we show that the SSEL approach can also be applied to investigate quantum phase transitions in 1D Ising and 1D XXZ models upon incorporating quantum sampling. We argue that the SSEL model simulates a special state function with higher-order correlations between physical quantities, and hence provides richer information than previous machine learning methods. Consequently, our SSEL method can be generally applied to the identification/classification of phase transitions even without explicit knowledge of the underlying theoretical models.
\end{abstract}

\maketitle

\section{Introduction}
The integration of machine learning (ML) techniques into theoretical and experimental physics has attracted substantial interest in recent years. A notable advantage of these methods, specifically supervised learning, lies in the efficient simulation of functional relationships between input data and output labels without requiring prior knowledge of the underlying theories or mechanisms \cite{intro_SL_1,intro_SL_2}. This opens a new research methodology to bridge the standard theoretical, experimental and numerical approaches to find new physics that cannot be observed before \cite{review_ML}.

One of the most extensively applied subjects is the identification of phase transition points through the observable data input \cite{SL_1, SL_2, SL_3, SL_4, SL_5, SL_6, SL_7,SL_8}. However, different from ordinary pattern recognition problems, where different objects are physically irrelevant with each other, the many-body phases in physics are actually based on the same Hamiltonian and connected by continuous system parameters. As a result, the predicted phase boundary may strongly depend on how the training and test regimes are selected in the continuous parameter space, and hence may be just an artifact if no theoretical guidance is available \cite{SSL_Ho}.

An alternative method, offered by unsupervised learning techniques, involves clustering input data within a reduced parameter space for the identification of distinct phases  \cite{USL_1,USL_2,USL_3, USL_4, USL_5, USL_6, USL_7, USL_8,USL_9,USL_10,USL_11,USL_12}. This approach hinges on the assumption that the data "distance" between different phases in the reduced parameter space is greater than the data fluctuation within each phase, facilitating the differentiation of various phases without the knowledge of their labels. Nevertheless, this assumption becomes tenuous near the phase boundary, where thermal or quantum fluctuations between distinct phases are substantial to render precise distinctions between data of different phases. Moreover, the estimated data "distance" is susceptible to the choice of algorithms and kernel functions, impeding the possibility to identify first-order, second-order, or other types of phase transitions.
\begin{figure}[ht]
    \centering
    \includegraphics[width=0.48\textwidth]{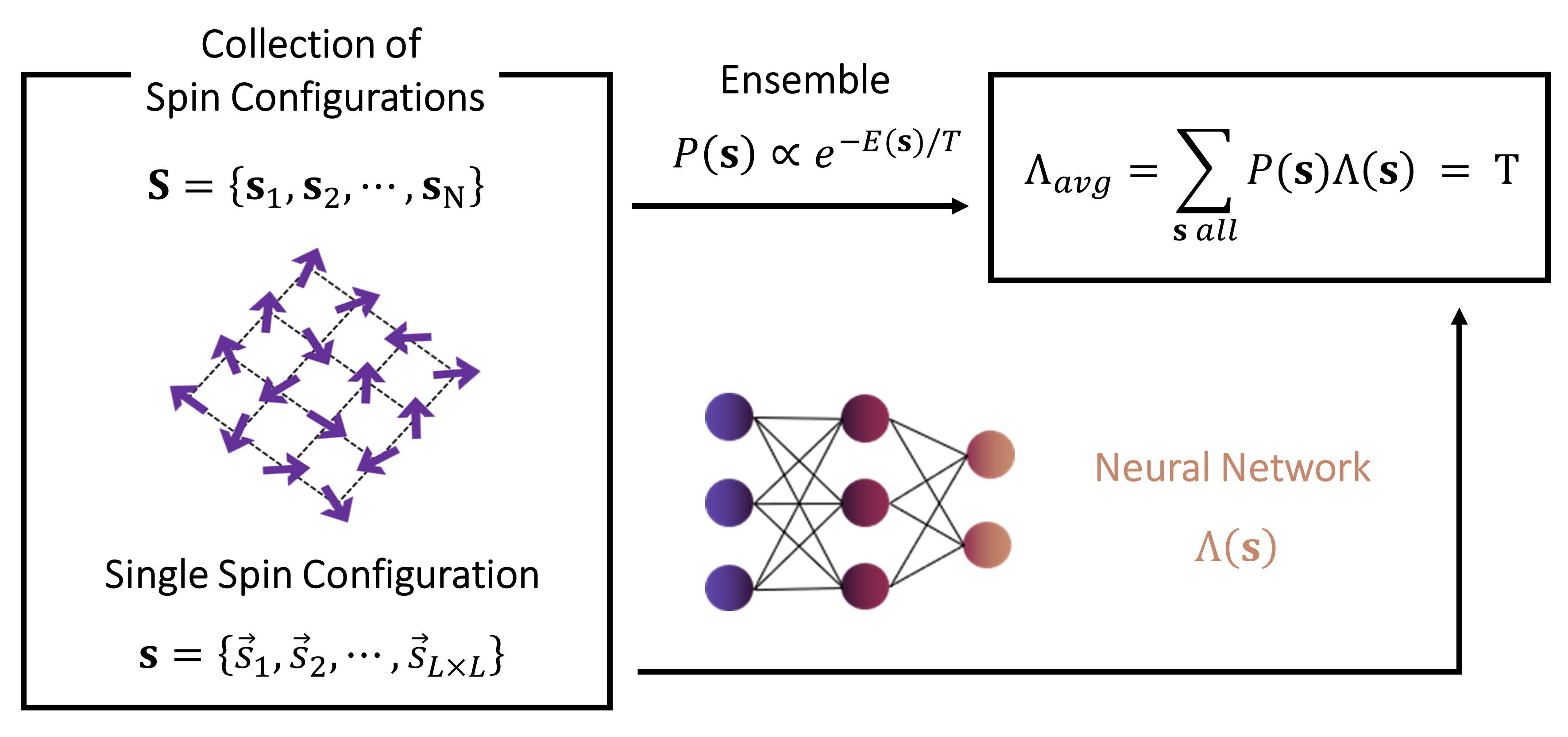}
    \caption{The scenario of the self-supervised ensemble learning (SSEL) method proposed in this paper. The neural network could be considered as a variational method to simulate the function (denoted by $\Lambda$) between a spin configuration (denoted by $\bf{s}$) and the system parameter (here is temperature $T$ for classical phase transition). $\Lambda_{avg}$ is obtained by ensemble averaging the neural network outputs of many spin configurations with a probability distribution, $P(\bfs)$, see Eq. (\ref{eqn:relation}).}
    \label{fig: intro_ML}
\end{figure}

In this work, we propose a self-supervised ensemble learning method (SSEL) to identify a variety of phase transitions in both classical and quantum many-body systems. Specifically, we employ the 2D Potts model and 2D Clock model as benchmarks for classical transitions and utilize the 1D spin-1/2 Ising model with a transverse field and 1D spin-1/2 XXZ model as benchmarks for quantum transitions. By generating sampled spin configurations and ensemble-averaging machine learning outputs, we construct a special "state function" (called state function with higher order correlation, SFHOC, see below) by associating the in-situ spin configurations and the known system parameters (e.g., temperature or external field) through the self-supervised learning approach. The phase transition point is then robustly determined by the position at which the fluctuations exhibit the most significant change as the system parameter varies. Through a systematic examination of fluctuation patterns in proximity to the critical points, we further demonstrate that  first-order, second-order, and Berezinskii-Kosterlitz-Thouless (BKT) phase transitions \cite{BKT_1,BKT_2,BKT_3,BKT_4} can be distinguished easily. We further discuss why our SSEL could simulate a SFHOC and hence could serve as a universal approach for the identification/classification of unknown phase transitions in matters, even without prior knowledge of the underlying theoretical mechanisms.

This paper is organized as follows: We introduce the scenario of the SSEL in Sec. \ref{sec:ssel} and use the 2D classical Ising model as an example to present how to identify a phase transition in Sec. \ref{sec:ising}. In Sec. \ref{sec:classical}, we consider the 2D Potts model and the 2D clock model to show how different phase transitions could be classified by the SSEL. In Sec. \ref{sec:quantum}, we consider the 1D spin-1/2 quantum Ising model and the 1D spin-1/2 XXZ model to demonstrate how the SSEL framework could be also applied to classify quantum phase transitions. Finally, in Sec. \ref{sec:discussion}, we argue the relationship between our SSEL method and previous methods and why it simulates a SFHOC for the identification of phase transitions. We then conclude this paper in Sec. \ref{sec:summary}. The comprehensive details of the implementation of our SSEL are provided in Appendix \ref{sec:appendix}.

\section{Self-Supervised Ensemble Learning}
\label{sec:ssel}

To explain the central concept of the SSEL model, we take a classical spin system as an illustrative example, see Fig. \ref{fig: intro_ML}. The system can be described by a collection of microscopic states via different spin configurations, $\bfs\equiv\{\vecs_1,\vecs_2,\cdots,\vecs_{L\times L}\}$ with $\vecs_i$ being the spin orientation at the $i$th site in a 2D system of size $L$. According to the basic assumption of statistical mechanics \cite{Andersonbook}, a macroscopic system could contain numerous microscopic states, each of which has an equal probability of being observed in a micro-canonical ensemble. When the system is in contact with a thermal reservoir at temperature $T$, the equilibrium probability distribution of these microscopic states reveals a Boltzmann distribution, $P(\bfs)\propto e^{-E(\bfs)/T}$, by balancing the system energy and its total entropy at the same time. Here $E(\bfs)$ is the system total energy of a given spin configuration $\bfs$.

As a result, there should be \textit{no} one-to-one functional relationship between the spin configuration of a microscopic state and any macroscopic physical quantity (say the temperature), because the same spin configuration can be observed at two different temperatures just with different probabilities. Therefore, one should not expect any machine learning approach to simulate the functional relationship between in-situ measurements (as the input feature) and a macroscopic quantity. 

On the other hand, from an experimental point of view, a physical quantity, here denoted by $x$, is actually identified by averaging many in-situ measurements of the same macroscopic state at the same temperature (or the same quantum state at zero temperature). The relationship between the averaged macroscopic physical quantities (e.g. $\langle x_i\rangle$, where $\langle\cdots\rangle$ denotes the ensemble average and $i=1,2,\cdots$ are the index of different quantities) could then be defined by a state function, $F(\{\langle x_i\rangle \})=0$) \cite{Pathriabook,Swendsenbook}. This traditionally defined state function could be in principle simulated by a machine learning model through a self-supervised learning approach and hence, as proposed by us before \cite{SSL_Ho}. 

However, inspired by the statistical nature of the experimental data via in-situ measurements (as described above), in the SSEL method we propose here, we numerically average the neural network outputs directly and try to associate it with a macroscopic physical quantity, say temperature, through the training process. More precisely, we have $N$ spin configurations, ${\bfS}\equiv\{\bfs_{1},...,\bfs_{N}\}$ ($N\gg 1$), to be sampled at a temperature $T$ according to the Boltzmann distribution, $P(\bfs)\propto e^{-E(\bfs)/T}$ through Monte Carlo simulation. These collections of spin configurations are subsequently fed into a neural network through a data-driven self-supervised learning approach, see Fig. \ref{fig: structure_ML}, and obtain the following averaged result for a given sampling, $\bfS$,
\begin{eqnarray}
\Lambda_{avg}(\bfS,T)=\frac{1}{N}\sum_{\bfs \in \bfS}\Lambda(\bfs)
\propto
\sum_{\text{all }\bfs} e^{-E(\bfs)/T} \Lambda(\bfs), 
\label{eqn:relation}
\end{eqnarray}
where $\Lambda(\bf{s})$ is the neural network outcome for a spin configuration as an input feature. In Sec. \ref{sec:discussion}, we will further discuss how this SSEL method is to simulate high-order correlations between physical quantities in a special state function, which has not been discussed before.

In order to successfully simulate the functional relationship between the collection of spin configurations and the temperature, we use the mean squared error (MSE) loss function as an estimator to minimize the error between averaged output and temperature:
\begin{equation}
L({\widetilde{\bfS}_1},...,{\widetilde{\bfS}_M})=\frac{1}{M}\sum_{m=1}^{M}|\Lambda_{avg}({\widetilde{\bfS}}_m,T_m)-T_m|^2,
\label{eqn: loss_function}
\end{equation}
where $\widetilde{\bfS}_m$ is the $m$-th input configurations generated at temperature $T_m$ and $M$=16 is the batch size (see the comprehensive details for the training method in Appendix \ref{sec:appendix}).
Note that this loss function ensures that the neural network can simulate the functional relationship between multiple configurations in each training batch and the unique macroscopic physical quantity (temperature here). This is the major difference between our SSEL method and previous methods, which uses averaged results (instead of in-situ measurements) as the input features \cite{SSL_Ho}.

After the training process, we will analyze the obtained standard deviation, $\Lambda_{std}(T)$, for the study of phase transition properties, where 
\begin{equation}
\Lambda_{std}(T) \equiv \left[\frac{1}{N}\sum_{{\bf s}\in{\bfS}}|\Lambda({\bf s})-\Lambda_{avg}(\bfS,T)|^2\right]^{1/2}.
\label{eqn:Lambda_std}
\end{equation}
Note that the sample dependence of $\Lambda_{std}(T)$ can be ignored if $N\gg 1$ (we take $N=2000$ throughout our calculation below). We also emphasize that in the SSEL we propose here, there is no test data needed and the fluctuations are simulated during the training process directly. Besides, the output label of the training process is simply the temperature or any experimentally controllable/known physical quantities, and therefore \textit{no} information about the phase transition (no matter if it exists or not) is needed either. As a result, this approach can be also applied in a quite general situation with different systems of matter.

\section{Example: 2D Ising Model} 
\label{sec:ising}
\begin{figure}[ht]
    \centering
    \includegraphics[width=0.48\textwidth]{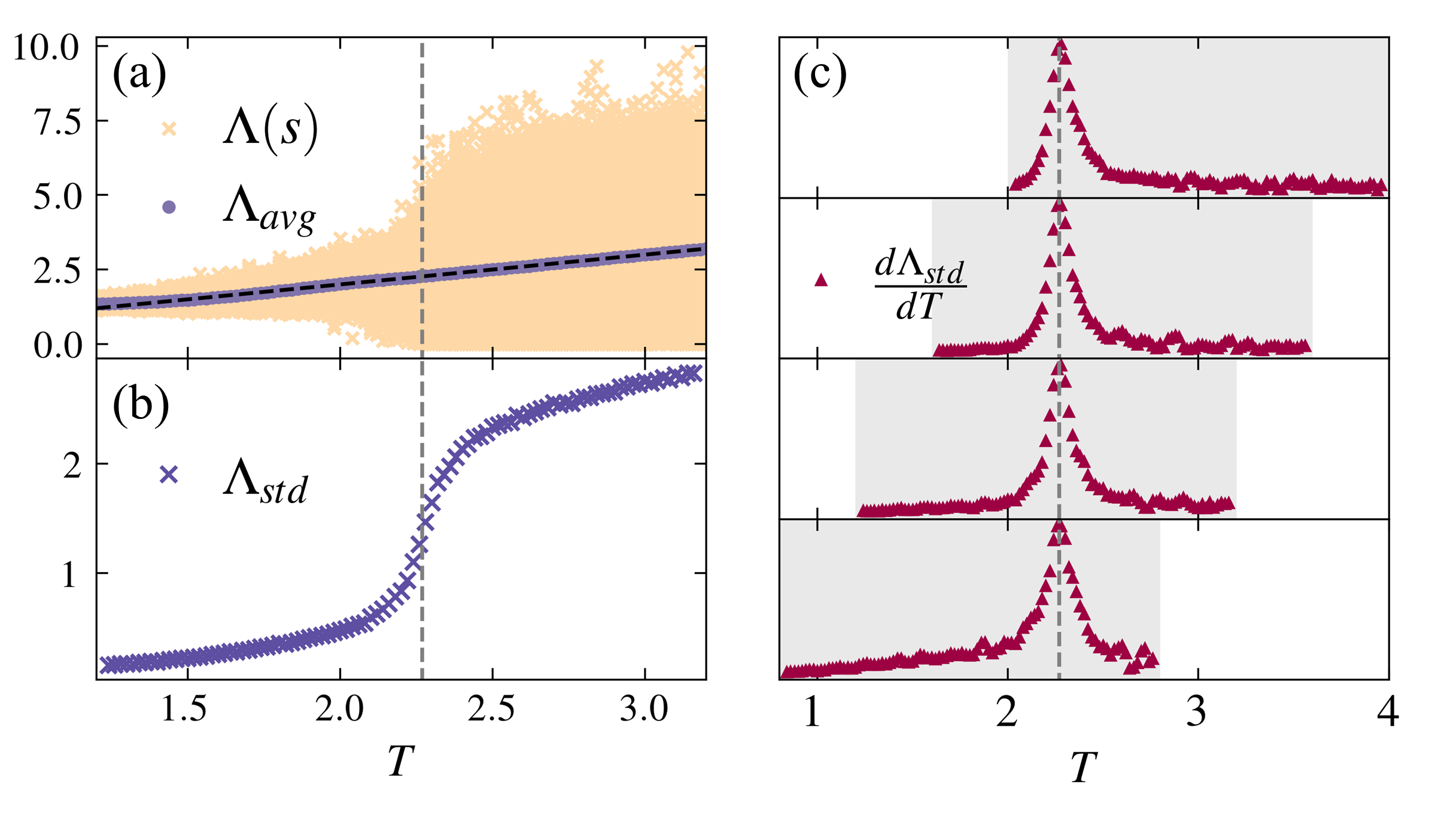}
    \caption{The SSEL results for a classical 2D Ising model with $L=48$. The vertical dashed line denotes the critical temperature $T_c=2.269$. (a) The individual outputs $\Lambda(\bf{s})$ (yellow dots) from the model for a given temperature and spin configuration. Their averaged value ($\Lambda_{avg}$) must be equal to the temperature (dashed line) due to the constraint of our loss function, see Eq. (\ref{eqn: loss_function}). (b) The calculated standard deviation of the model outputs ($\Lambda_{std}$) which is defined in Eq. (\ref{eqn:Lambda_std}). (c) The first temperature derivative of $\Lambda_{std}$ has a significant peak, which is pinned almost at the same point near $T_c$, and is almost the same for different training regimes (gray region).}
    \label{fig: scheme_example}
\end{figure}

We first consider the classical Ising model on a 2D square lattice for the demonstration. The system energy is given by $E = -\sum_{<i,j>}s_{i}s_{j}$, where $s_{i}\in\{1,-1\}$. We sample $N=2000$ configurations at each temperature  and display all the outputs for each spin configuration as shown in Fig. \ref{fig: scheme_example}(a). It should be noted that even the average output ($\Lambda_{avg}$) is required by the loss function to be equal to the temperature (see Eq. (\ref{eqn: loss_function})), the individual output could be any value without strict constraint. In fact, its distribution becomes wider as the temperature increases, reflecting the broadened Boltzmann weight at higher temperatures. 

In Fig. \ref{fig: scheme_example}(b) and (c), we present the calculated $\Lambda_{std}(T)$ and its temperature derivative. The former is a monotonically increasing function as anticipated, while the latter exhibits a significant peak. Such a non-trivial peak structure is not an artifact of machine learning, as its position remains robustly pinned at the same point, independent of the temperature regions we considered (see the gray area in (c)). The maximum position of $d\Lambda_{std}/dT$ aligns well with the known critical temperature $T_c\sim2.269$ \cite{Isingbook} (vertical dashed lines), thus serving as a reliable indicator of the phase transition. On the other hand, if no phase transition exists within the temperature range of our training data, the calculated $\Lambda_{std}$ would simply increase monotonically without the sign change in curvature, i.e. no peak structure in its derivative. 

The physics underlying the aforementioned example can be understood in the following simple scenario: First, it is known that a phase transition is mathematically defined by the singularity of the system state function \cite{Pathriabook}, which represents a functional relationship between various macroscopic physical quantities ($\langle x_i\rangle$, obtained after the ensemble average of a macroscopic state). Since spin configurations are sampled by the system temperature through Boltzmann distribution, it is reasonable to expect that the SSEL model attempts to simulate a specific "state function", i.e. $\Lambda_{avg}(\bfS,T)=T$ (see Eqs. (\ref{eqn:relation}) and (\ref{eqn: loss_function})). Consequently, near the critical point of two competing phases, the fluctuations in the simulated results show the most significant increase, in addition to the ordinary thermal fluctuations. In Sec. \ref{sec:discussion}, we will argue why this kind of special "state function" may include higher order correlations of physical quantities and hence can be applied for the identification and classification of various phase transitions.

In the rest of this paper, we will present more examples and analyze their properties based on the SSEL results. 

\begin{figure}[ht]
    \centering
    \includegraphics[width=0.48\textwidth]{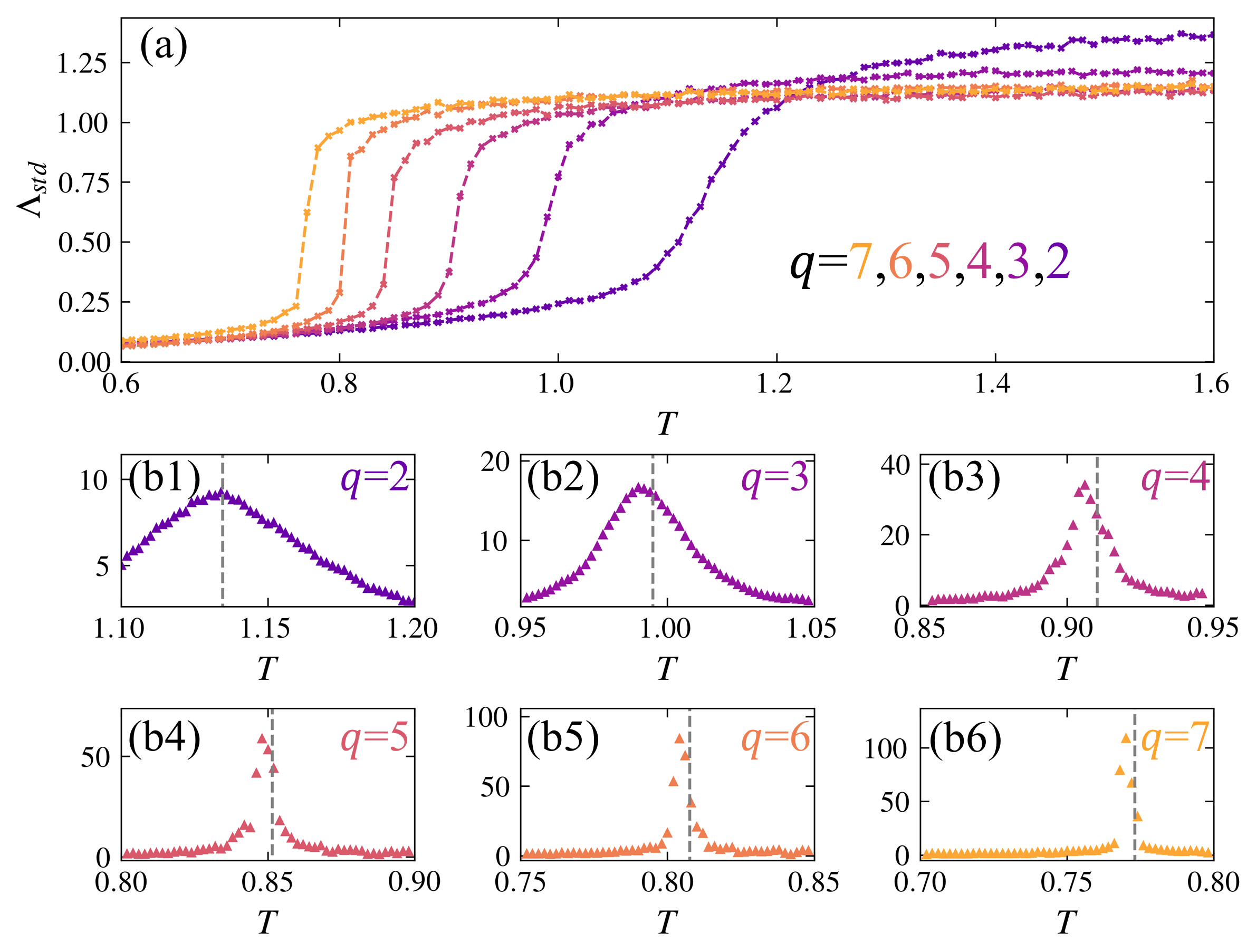}
    \caption{The SSEL results for a Potts model. (a) The calculated $\Lambda_{std}$ with $L$=48 for $q=2\sim7$. (b1)-(b6) are the temperature derivatives of  $\Lambda_{std}$ for $q=2\sim7$ within their critical temperature (vertical dashed line), respectively. They are computed by feeding more configurations to the well-trained SSEL model we showed in (a).}
    \label{fig: potts}
\end{figure}
\section{Classify Finite Temperature Phase Transitions} 
\label{sec:classical}

We proceed to examine the Potts model \cite{Potts, Potts_review} and the Clock model, also known as the planar Potts model \cite{Clock, Potts_review, Clock_critical_2}, as benchmarks to investigate the applicability of the SSEL model not only for pinpointing $T_c$ but also for classifying different phase transitions. In contrast to the 2D Ising model discussed above, the Potts model and the Clock model permit a much richer variety of spin configurations, resulting in much richer phase diagrams and diverse phase transitions. More precisely, in these two models, each spin can adopt $q\in N$ possible orientations or states, i.e., $\theta_{r_i} = 2\pi r_i/q$, where $\theta_{r_i}$ represents the orientation of the spin in the 2D plane, determined by its state index, $r_i\in{1,2,\cdots,q}$, at the $i$th site.

We start our analysis with the Potts model, in which neighboring spins interact only when they have the same orientation:
\begin{equation}
    E_{Potts} = -\sum_{\langle i,j\rangle} \delta(\theta_{r_i},\theta_{r_j}),
    \label{eqn:Potts model}
\end{equation}
where $\delta$ is the Kronecker delta. The phase transition is known to be second-order for $1<q\leq 4$ \cite{Potts_critical_1} and first-order for $q>4$ \cite{Potts_critical_2}, with the critical temperature $T_c =1/\ln(1+\sqrt{q})$ \cite{Potts_critical_3}. Following the same procedure as the 2D Ising model, we provide the standard deviation of the outputs ($\Lambda_{std}(T)$) for q=$2\dots 7$ with $L=48$ in Fig. \ref{fig: potts}(a). One could find that $\Lambda_{std}$ exhibits a steeper slope as $q$ increases. Additionally, in Fig. \ref{fig: potts}(b1)-(b6), we present the calculated $d\Lambda_{std}/dT$ as a function of temperature $T$ for different $q$ values. It can be observed that $d\Lambda_{std}/dT$ reaches its maximum near the known phase transition points (vertical dashed lines) in all of these models, while their patterns appear quite distinct: the "peak" becomes much sharper for $q>4$, indicating possible different critical behaviors near $T_c$ compared to those for $q\leq 4$. 

To quantitatively examine the possibility of distinguishing first-order and second-order phase transitions in the 2D Potts model via our SSEL method, we must consider the finite size scaling, as the true phase transition and the singularity of the state function are well-defined only in the limit of an infinite-size system ($L\to\infty$). As a result, we propose a phenomenological approach to analyze the continuity of $\Lambda_{std}$ by assuming its general form to be the following function:
\begin{equation}
    \Lambda_{std}(T)=\text{sgn}(T-T_c)A|T-T_c|^{\gamma}+B,
    \label{eqn:gamma_def}
\end{equation}
where $\text{sgn}(T-T_c)$ is the sign function. $\gamma$ is the associate exponent obtained by fitting the numerical results of our SSEL. $A$ and $B$ are the other two fitting parameters, and $T_c$ is determined by the peak position of $d\Lambda_{std}/dT$. We note that $\gamma$ is in general positive and smaller than unity from the numerical results shown in Fig. \ref{fig: potts}(a).

\begin{figure}[ht]
    \centering
    \includegraphics[width=0.48\textwidth]{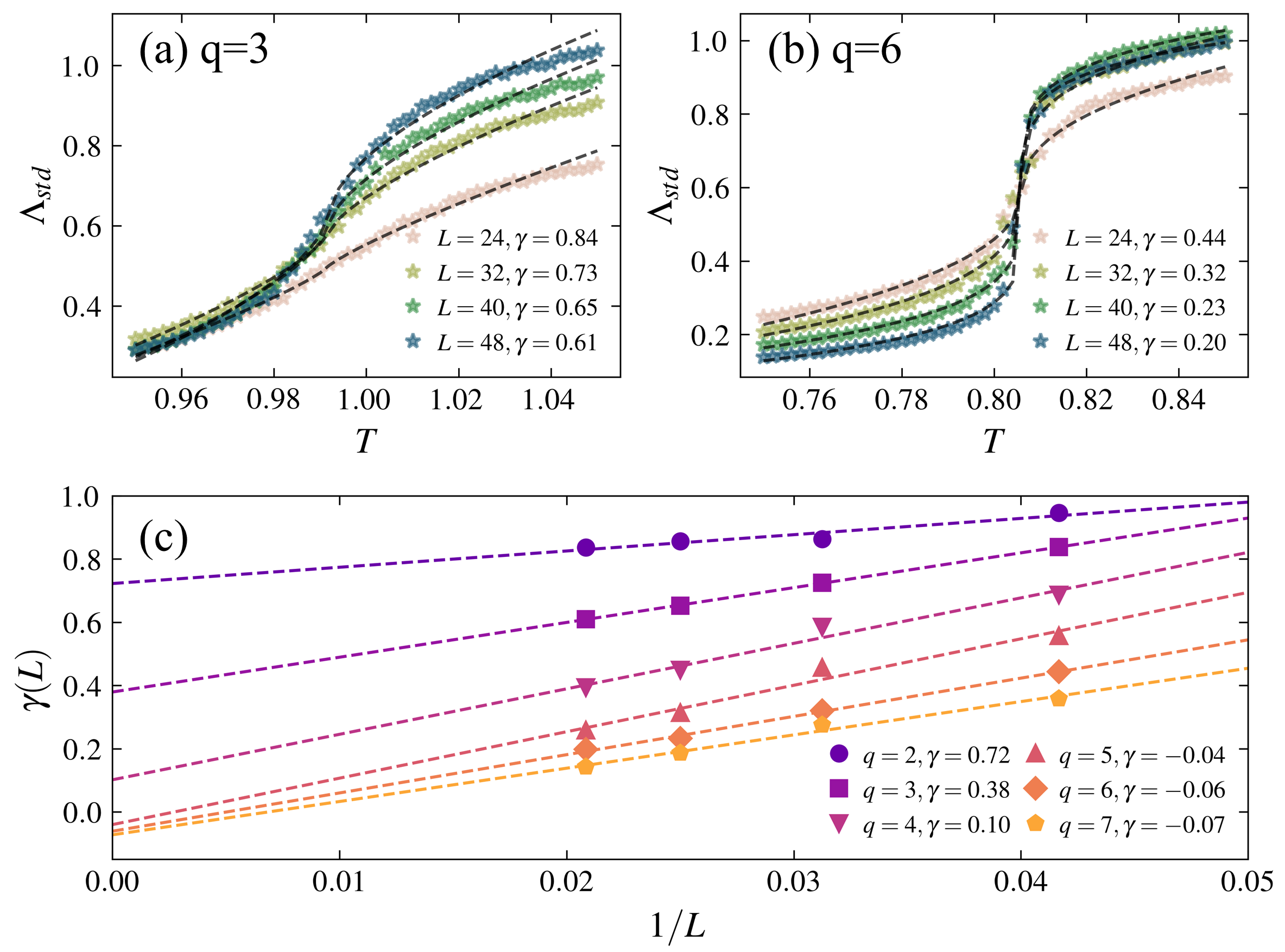}
    \caption{The estimation of the exponent ($\gamma$) for a Potts model. In (a) and (b), we plot the calculated $\Lambda_{std}$ (dots) and its fitting result (dashed line) with $L=24,32,40,48$ for $q=3$ and $6$, respectively. (c) depicts the extrapolation of the estimated exponents ($\gamma$) to the thermodynamic limit, where it turns out positive for $q=2\sim4$ and negative for $q=5\sim7$.}
    \label{fig: potts_scaling}
\end{figure}

In Fig. \ref{fig: potts_scaling}(a) and (b), we present several fitting results of $\Lambda_{std}(T)$ for different system sizes with $q=3$ and $q=6$, respectively. Furthermore, in Fig. \ref{fig: potts_scaling}(c), we display the scaling behavior of these fitting exponents for various system sizes and extrapolate them to the infinite-size limit. The estimated exponents at $L=\infty$ are found to be positive for $q\leq 4$, while they become negative for $q>4$, implying the possible existence of a finite step discontinuity of $\Lambda_{std}(T)$ in the thermodynamic limit. This finding is consistent with known results that the phase transition is second-order for $q\leq 4$ and becomes first-order for $q>4$. 

We note that the phenomenological analysis provided above can also be applied to other models directly, indicating that our SSEL method could not only detect the presence of phase transitions but also classify them based on their critical behaviors through finite-size scaling. This capability represents one of the most significant advantages of our SSEL approach, setting it apart from other supervised or unsupervised learning techniques. 

To further investigate the potential of our SSEL model in differentiating the BKT transition and the second-order phase transition, we consider the 2D Clock model with the following total energy:
\begin{equation}
E_{Clock} = -\sum_{\langle i,j\rangle} \cos(\theta_{r_i} - \theta_{r_j})
\end{equation}
It has been shown that the phase transition of a 2D Clock model is second-order for $q\leq 4$ and becomes BKT type for $q>4$ \cite{Clock_critical_1,Clock_critical_2}. It becomes the standard 2D $XY$ model in the limit of $q\to\infty$ \cite{XY}.

\begin{figure}[ht]
    \centering
    \includegraphics[width=0.48\textwidth]{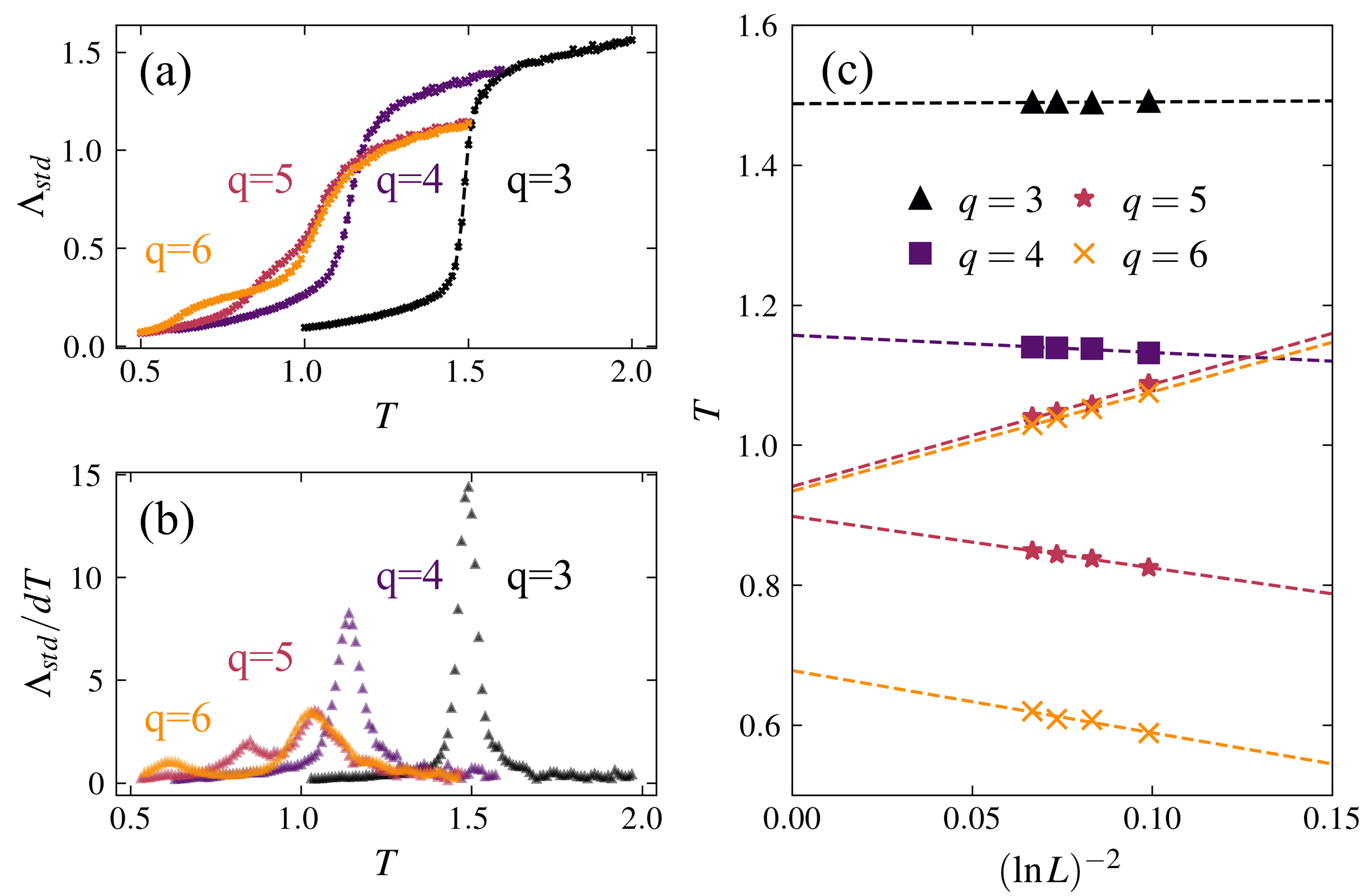}
    \caption{The SSEL results for a Clock model. In (a) and (b), we show the calculated $\Lambda_{std}$ and its temperature derivative for $q=3$, 4, 5, 6 with $L$=48, respectively. The double-peak structure observed for $q=5$ and 6 in (b) indicates the presence of two BKT transitions. (c) shows the finite-size scaling extrapolation of the critical temperatures, corresponding to the peak positions in panel (b). The finite-size dependences reveal the significant difference between a second-order phase transition ($q\leq 4$) and a BKT transition ($q>4$).}
    \label{fig: clock}
\end{figure}

In Fig. \ref{fig: clock}(a) and (b), we display the standard deviation of the model outputs ($\Lambda_{std}$) and their temperature derivatives for $q=3$, 4, 5, 6 with $L$=48, respectively. One can clearly observe the dramatic differences between the results for $q\leq 4$ and those for $q>4$, where the latter exhibits a double-peak structure in $d\Lambda_{std}/dT$, indicating two phase transitions within the temperature regime we calculated. 

In Fig.\ref{fig: clock}(c), we show how the values of critical temperatures, obtained by the peak positions of $d\Lambda_{std}/dT$, vary as a function of system size $L$. Notably, the results for $q>4$ demonstrate a much stronger system size dependence, scaled with $\ln(L)^{-2}$, in contrast to the nearly constant values for $q\leq 4$. This observation aligns with the known theoretical understanding that the 2D Clock model exhibits BKT transitions for $q>4$, wherein critical temperatures are highly sensitive to system size due to the power-law scaling of the correlation function \cite{BKT_1,BKT_2,BKT_3,BKT_4}. Moreover, by conducting the scaling analysis and extrapolating the peak positions, we determine the critical temperatures in the thermodynamic limit to be $0.898$ and $0.941$ for $q=5$, $0.678$ and $0.934$ for $q=6$, respectively. These values are in good agreement with the results from tensor network simulations ($0.9059$ and $0.9521$ for $q=5$, $0.6901$ and $0.9127$ for $q=6$) \cite{Clock_critical_2}. As a result, we can conclude that our SSEL method can be also employed to distinguish the BKT transition from the regular second-order phase transition by analyzing how the peak positions of $d\Lambda_{std}/dT$ change as a function of the system size.

\section{Quantum spin model} 
\label{sec:quantum}
Finally, we examine the 1D Ising model with a transverse field (IMTF) and 1D spin-1/2 XXZ model as benchmarks to demonstrate how the SSEL model can be extended to identify the quantum transitions. Different from the finite temperature cases, the probability to sample a spin configuration should be derived from the ground state wavefunction according to quantum mechanics, i.e. $P(\bfs)\propto |\langle \bfs|G\rangle|^2$, where $|G\rangle$ is the ground state and $|\bfs\rangle$ is a basis vector with the spin configuration $\bfs$. The output label can be simply the external parameter in Hamiltonian, such as the external field in our current cases.

\begin{figure}[ht]
    \centering
    \includegraphics[width=0.48\textwidth]{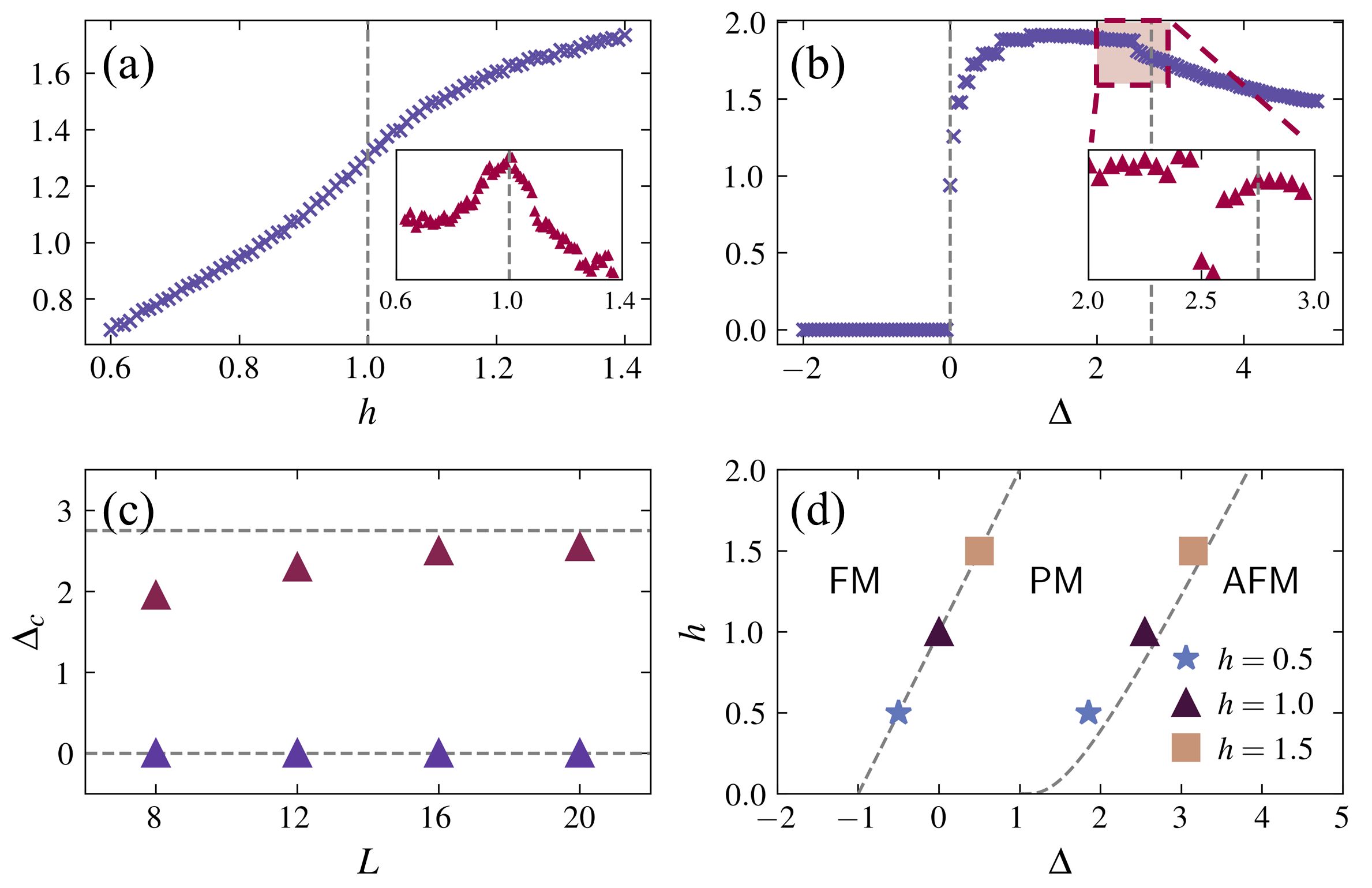}
    \caption{
    The calculated $\Lambda_{std}$ and its first derivative (inset) for (a) 1D IMTF model with $L$=24 and (b) 1D spin-1/2 XXZ model with $L$=20 and  $h=1$. (c) The finite size dependence of the critical anisotropicity, $\Delta_c$, for 1D XXZ model at $h=1$. (d) The calculated quantum phase diagram for the 1D XXZ model with $h$=0.5, 1.0, and 1.5. Dashed lines in all these panels are the theoretical values of the critical points in the thermodynamic limit ($L\to\infty$).
    }
    \label{fig: quantum}
\end{figure}

The Hamiltonians of 1D IMTF and 1D XXZ are known to be the following:
\begin{eqnarray}
\hat{H}_{\text{IMTF}} &=& -\sum_{i}\left(
\hatsigma^{z}_{i}\hatsigma^{z}_{i+1}
+h\hatsigma^{x}_{i}\right) \\
\label{eq:IMTF}
H_{\text{XXZ}} &=& \sum_{i}(
\hatsigma^{+}_{i}\hatsigma^{-}_{i+1}
+\hatsigma^{-}_{i}\hatsigma^{+}_{i+1}
+\frac{\Delta}{2}\hatsigma^{z}_{i}\hatsigma^{z}_{i+1}
-h\hatsigma^{z}_{i})
\label{eq:XXZ}
\end{eqnarray}
where $\hatsigma^{x,y,z,\pm}_{i}$ is Pauli spin operator as regular definition, $h$ is the external field and $\Delta$ is anisotropy coupling.  In Fig. \ref{fig: quantum}(a), we display the change in $\Lambda_{std}$ calculated for IMTF as a function of $h$ (with an inset depicting its derivative). One could find that the quantum fluctuation increases from the ordered ferromagnetic phase (smaller $h$) to the disordered paramagnetic phase (larger $h$), while its derivative also exhibits a peak near the critical point $h=1$. 
In Fig. \ref{fig: quantum}(b), we show the calculated $\Lambda_{std}$ with its derivative in the inset for the results of the 1D XXZ model as a function of its anisotropicity with $h=1$. It is clear that $\Lambda_{std}$ has \textit{zero} fluctuation for $\Delta<0$, indicating strong ferromagnetism (FM), while the fluctuation starts to exhibit a sudden jump as $h=0^+$. It suggests the presence of a first-order transition from the fully polarized state to the magnon excited state (compared to the Potts model for $q=6$ in Fig. \ref{fig: potts_scaling}(b)). On the other hand, the fluctuations become decreasing in much larger $\Delta$, because the ground state tends to only two degenerate N{\'e}el states and shows antiferromagnetism (AFM). The derivative of $\Lambda_{std}$ also exhibits a local minimum near $\Delta=2.75$, implying the position of a quantum phase transition. In other words, between the FM state for $\Delta<0$ and the AFM state for $\Delta > 2.75$, the ground state corresponds to paramagnetism (PM). The transition between PM and AFM is also known as a BKT transition at zero temperature \cite{XXZbook}.

In Fig \ref{fig: quantum}(c), we show how the obtained critical anisotropicity ($\Delta_c$) changes as a function of the system size $L$ for $h=1$. It is clear to see that there is no change at all for the first order phase transition at $\Delta=0$, while $\Delta_c$ exhibits significant changes as $L$ increases and approaches its theoretical value in the thermodynamic limit as $L>20$. Such a strong system size dependence reflects the key characteristic of BKT transition. In Fig. \ref{fig: quantum}(d), we present the SSEL prediction of the two critical anisotropicities with $L=20$ and $h=0.5, 1.0, 1.5$. It can be seen that these finite-size results are well close to the theoretical values in the thermodynamic limit, especially in the strong-field regime. As a result, it is reasonable to believe that the SSEL framework proposed in this paper can be also applied to quantum many-body systems.

\section{Discussion} 
\label{sec:discussion}
It is instructive to compare the SSEL method proposed here to our previously self-supervised learning (SSL) model \cite{SSL_Ho}. We have to emphasize that the SSL can also be applied to identify phase transitions, but cannot classify their properties or types as our SSEL model here. The fundamental difference between these two self-supervised approaches is their input features: it is from in-situ measurements in our SSEL, while it is a certain physical quantity obtained \textit{after the ensemble average} in the SSL. The former (in-situ measurements) certainly contains much more fundamental information about the state of the matter than the latter, making the SSEL possible to identify phase transitions even in the training regime (see Fig. \ref{fig: scheme_example}) via the fluctuations of the model output. 


More precisely speaking, in the training regime of the SSL method, the machine learning model could always simulate the regular state function very well, which is the functional relationship between the input and output physical quantities, obtained after the ensemble average. As a result, the fluctuations of the model output obtained in the test regime, due to the mismatch between the trained state function and the exact one, can not contain sufficient statistical information to classify the phase transition nature. In contrast, the statistical properties of in-situ measurements have already been included in the SSEL method as shown in Eq. (\ref{eqn:Lambda_std}). This is why our SSEL could distinguish different types of phase transitions, as demonstrated in this work.

On the other hand, we know the average of a function of variables is in general different from the function of averaged variables, i.e. $\langle F(\{x_i\})\rangle\neq F(\{\langle x_i\rangle\})$. Their differences clearly are resulted from the higher-order correlation functions between these physical quantities, say $\langle x_i x_j\rangle$. As a result, it is reasonable to expect that our SSEL method can be considered to be a very general method to simulate a state function with higher-order correlations (SFHOC), say 
$\langle F(\{x_i\})\rangle=0$, which contains more information than regular state function, $F(\{\langle x_i\rangle\})=0$. However, we admit that it is still not clear how to derive the scaling exponents (say $\gamma$ in Eq. (\ref{eqn:gamma_def})) from existing many-body or statistical theories. 

The advantage of the SSEL can be also appreciated from the experimental point of view because these in-situ data can be obtained more easily since the finite temperature and/or quantum fluctuation effects have been included directly through multiple measurements. These fluctuations in a realistic experiment can be included in our SSEL method through machine learning, making it possible to analyze the phase transition properties more directly. The fundamental reason is that the SSEL method is to directly simulate the functional relationship between fluctuating in-situ measurements and the physical quantities through a SFHOC and hence provides an independent approach to investigate the phase transition properties. 

\section{Summary}
\label{sec:summary}

In summary, we have provided a self-supervised ensemble learning method for effectively identifying classical and quantum phase transitions. By developing a neural network to simulate the state function of matter, our method demonstrates the capability to explore the presence of critical phenomena and classify different types of phase transitions, including first-order, second-order, and BKT transitions. The SSEL approach does not rely on any specific assumption about the underlying theoretical models and hence offers an \textit{independent} methodology of extracting valuable information about the physical system solely from experimental measurements. We argue the SSEL method is to simulate the stat function of higher-order correlations between physical quantities and hence contains more information than previous machine learning methods for the classification of phase transitions. Further investigation of the theoretical understanding and its application to other many-body properties will be continued in future studies.

\section{Acknowledgement} 
We thank Kuan-Ting Chou, Chen-Yu Liu, Chung-Yu Mou, Yi-Ping Huang, Jhih-Shih You, Po-Yao Chang, and Po-Chung Chen for the valuable discussion. This work is supported by the National Center for Theoretical Sciences, the Higher Education Sprout Project funded by the Ministry of Science and Technology, and the Ministry of Education in Taiwan. DWW is supported under the grant MOST 110-2112-M-007-036-MY3.




\bibliography{main}

\appendix
\section{Model Structure and Hyper-Parameters for the Implementation of Self-Supervised Ensemble Learning (SSEL) Model}
\label{sec:appendix}

\begin{figure*}[ht]
    \centering
    \includegraphics[width=1.0\textwidth]{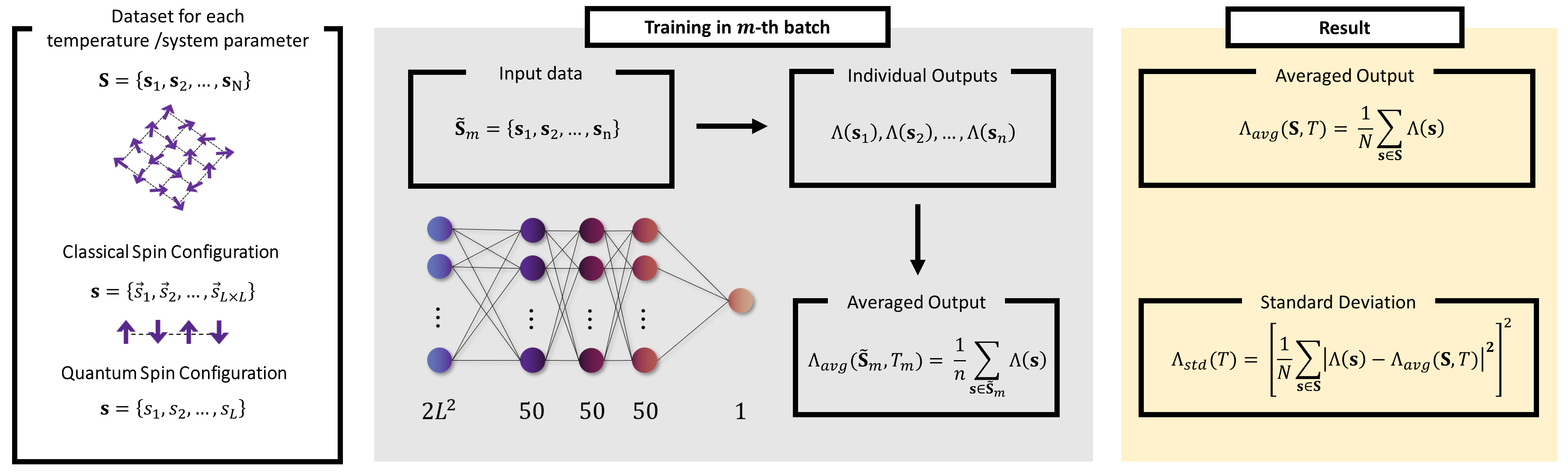}
    \caption{The scheme of the self-supervised ensemble learning (SSEL), taking a 2D classical spin system of size $L\times L$ as an example. In the gray (middle) panel, we show the neural network architecture and the input/output data structure for the SSEL method. The unit of input feature is $\widetilde{\bfS}$, which contains $n$=50 configurations. The fully-connected neural network has $2\times L^2$ nodes for the input layer, 50 nodes for the hidden layer, and only one node for the output layer. The activation function for each layer is SeLU \cite{SELU}, while the one in the output layer is ReLU in order to make the output positive. Here the batch size is $M=16$. In the yellow (right) panel, we show the formulas of ML results we presented in our manuscript. They are computed by using the whole collection $\bfS$, i.e. $N=2000$ configurations for each temperature.
    }
    \label{fig: structure_ML}
\end{figure*}

\begin{table}[ht] 
	\renewcommand{\arraystretch}{1.5}
	\tabcolsep=12pt   
	\centering
    \begin{tabular}{c c}
    \hline\hline
    Classical $(N=2000)$ & Temperature\\
    \hline
	2D Ising & $T = [1.2, 1.22,\cdots, 3.2]$\\ 
	2D Potts $(q=2\sim 7)$ & $T = [0.6, 0.61,\cdots, 1.6]$\\
	2D Clock $(q=3)$ & $T = [1.0, 1.01,\cdots, 2.0]$\\
        2D Clock $(q=4)$ & $T = [0.6, 0.61,\cdots, 1.6]$\\
        2D Clock $(q=5,6)$ & $T = [0.5, 0.51,\cdots, 1.5]$\\
    \hline\hline
    Quantum $(N=20000)$ & External parameter\\
    \hline
	1D IMTF & $h = [0.6, 0.61,\cdots, 1.4]$\\
	1D XXZ & $\Delta = [-2.00, -1.99,\cdots, 5.00]$\\
    \hline\hline
    \end{tabular}
    \caption{
    The sampled dataset we use for the training process in the manuscript. For each temperature/external parameter, we sample $N=2000$ spin configurations for the classical spin model and $N=20000$ spin basis for the quantum spin model. The model can be still well-trained if fewer training data are used. 
    }
    \label{tab:data}
\end{table}

In this appendix, we will introduce the model structure and hyper-parameters for the implementation of our SSEL model in the manuscript for both classical and quantum spin models. The training dataset we mentioned is listed in Tab. \ref{tab:data} as well. We generate the spin configurations of the 2D classical spin model by the Swendsen–Wang Monte Carlo algorithm \cite{SwendsenWang} and sample the spin basis of the 1D quantum spin model directly from its groundstate wavefunction, which is computed through the Lanczos algorithm \cite{Lanczos}.

In Fig. \ref{fig: structure_ML}, we show the neural network architecture for the classical spin system of size $L\times L$ as an example. The input layer has $2\times L^2$ nodes, corresponding to the planar orientations of $L\times L$ spin sites. More precisely, $\bfs\equiv\{\vecs_1,\vecs_2,\cdots,\vecs_{L\times L}\}$ = $\{s^x_1,s^y_1,\cdots,s^x_{L\times L},s^y_{L\times L}\}$, where $(s^{x}_{i},s^{y}_{i}) = (\cos\theta_{r_i}, \sin\theta_{r_i})$ is the spin orientation at $i$-th site in a 2D system of size $L$.  After connecting three consecutive fully-connected layers with 50 nodes with the SeLU activation function \cite{SELU}, we add an output layer with one node with the ReLU activation function in order to make the averaged output positive (i.e. temperature).

One should note that the sampling collection of microscopic configurations $\bfS$ at a specific temperature is not unique for different sampling trials, even though their macroscopic quantities are identical. Therefore, it is inappropriate to train SSEL to use only one possible $\bfS$ for each temperature, otherwise the over-fitting problem will lead to incorrect results. To properly simulate the functional relationship we are concerned with, we divide each collection $\bfS$, which contains $N=2000$ configurations, into 40 subsets $\widetilde{\bfS}$ to represent 40 different sampling trials. In other words, we have $N/n=40$ collections of configurations $\widetilde{\bfS}$ for each temperature, where $\widetilde{\bfS}$ contains $n=50$ configurations and is considered to be the unit of our input data.

Consequently, the $m$-th averaged output in a batch training is computed by the $n$=50 configurations in the given input data $\widetilde{\bfS}_m$, as shown in the gray panel of Fig.\ref{fig: structure_ML}. The corresponding loss function for the batch average of $M=16$ inputs is given as follows:
\begin{equation}
L({\widetilde{\bfS}_1},...,{\widetilde{\bfS}_M})=\frac{1}{M}\sum_{m=1}^{M}|\Lambda_{avg}({\widetilde{\bfS}}_m,T_m)-T_m|^2,
\label{eqn: loss_function_appendix}
\end{equation}
where $\widetilde{\bfS}_m$ is generated at temperature $T_m$.

We note that this division is only used in the training process. All the ML results presented in our manuscript, as illustrated in the yellow panel of Fig.\ref{fig: structure_ML}, are computed by using the whole configurations $\bfS$, i.e. $N$=2000 configurations for each temperature. 

On the other hand, for the quantum spin model of size $L$ in our manuscript, the number of configurations in a divided subset is $n=2000$, i.e. we have $N/n=10$ collections of configurations $\widetilde{\bfS}$ for each external parameter. We keep the same training scheme and neural network structure depicted in Fig.\ref{fig: structure_ML} but adjust the number of input nodes to $L$, i.e. $\bfs\equiv\{s_1,s_2,\cdots,s_L\}$ with $s_i\in\{1,-1\}$ being the spin state at $i$-th site in a 1D system of size $L$. 

We note that for quantum spin models, we do not apply the ReLU activation function in the output layer because there is no need to be positive for the output external parameter, especially for the negative anisotropy coupling in the 1D XXZ case.

\end{document}